\documentclass[prl,floatfix,twocolumn,showpacs,amsmath,amssymb]{revtex4}
\usepackage{graphicx}
\usepackage{dcolumn}
\usepackage{bm}

\begin{document}
\title{Jastrow theory of the Mott transition in bosonic Hubbard models}

\author{Manuela Capello,$^{1,2}$ Federico Becca,$^{1,2}$ Michele Fabrizio,$^{1,2,3}$
Sandro Sorella,$^{1,2}$}
\affiliation{
$^{1}$  International School for Advanced Studies (SISSA), Via Beirut 2-4, 
I-34014 Trieste, Italy \\
$^{2}$ CNR-INFM-Democritos National Simulation Centre, Trieste, Italy. \\
$^{3}$ International Centre for Theoretical Physics (ICTP), P.O. Box 586, I-34014 Trieste, Italy
}

\date{\today}

\begin{abstract}
We show that the Mott transition occurring in bosonic Hubbard models can be 
successfully described by a simple variational wave function that contains all important  
long-wavelength correlations. Within this approach, a smooth metal-insulator transition 
is made possible by means of a long-range Jastrow correlation term that binds in real space density  
fluctuations. We find that the Mott transition has similar properties 
in two and three dimensions but differs in one dimension. We argue that our description 
of the Mott transition in terms of a binding-unbinding transition is of general validity 
and could also be applied to realistic electronic systems.
\end{abstract}


\maketitle

Stimulated by the discovery of many strongly-correlated materials which, on the verge 
of becoming Mott insulators, display interesting and unusual properties, 
a huge theoretical effort has been devoted in the last decades to clarify the 
interaction-driven Mott metal-insulator transition (MIT).~\cite{mott} 
In spite of that, a full comprehension of this phenomenon is still lacking, 
even though, in the limit of infinite-coordination lattices,  
the whole dynamical behavior across the MIT can be uncovered
thanks to Dynamical Mean-Field Theory.~\cite{georges}  
As a matter of fact, the Mott phenomenon is not specific of fermions but 
also occurs in bosonic systems,~\cite{fisher} that have recently become popular 
in the context of optical lattices, where a MIT can be actually realized 
experimentally.~\cite{greiner} 

The prototypical Hamiltonian to describe a MIT both for fermions and bosons 
is the Hubbard model 
\begin{equation}\label{hambose}
{\cal H} = - \sum_{ij,\,\sigma}\, 
\bigg(t_{ij}\,b^\dagger_{i\sigma} b^{\phantom{\dagger}}_{j\sigma} + H.c.\bigg)
+ \frac{U}{2} \sum_i \, n_i\, (n_i-1),
\end{equation}
where $b^\dagger_{i\sigma}$ ($b_{i\sigma}$) creates (annihilates) a particle at site $i$ 
with spin $\sigma=-S,\dots,S$, being $S$ half-odd-integer for fermions and integer 
for bosons and $n_i=\sum_\sigma\,b^\dagger_{i\sigma} 
b^{\phantom{\dagger}}_{i\sigma}$ the local density operator. 
The Hubbard model~(\ref{hambose}) at commensurate densities generally shows two different 
phases: If $U<U_c$  particles are delocalized, which implies a metallic behavior 
for fermions (unless a Stoner instability leads to magnetically ordered phases) 
and superfluidity for bosons; instead, when $U>U_c$ the model describes 
a Mott insulator where coherent motion is suppressed. Presumably, for any $S\not = 0$, the Mott 
insulating phase is accompanied by translational and eventually spin-rotational symmetry 
breaking. However, the latter is merely a consequence of the Mott phenomenon and should 
not be identified as the driving mechanism leading to the insulating behavior, 
which arises from a strong suppression of charge fluctuations. Indeed, a MIT does occur also 
for $S=0$ bosons, in which case no symmetry breaking is expected within the Mott 
insulator. 
In this work we focus on the bosonic Hubbard model by describing the Mott phase with a 
strongly correlated variational wave function. 
An advantage of considering bosons is that we can directly compare 
the variational outcome with numerically exact results obtained by Green's function 
Monte Carlo (GFMC).~\cite{nandini,calandra} 
Therefore, by means of this comparison, we can establish the key ingredients that must
be included in the variational state for a faithful representation of a
genuine Mott insulating state.

In spite of the fact that the variational approach is a simple and well established 
technique, its application to the MIT turns out to be extremely difficult. 
For instance, the celebrated Gutzwiller wave function 
$|\Psi_G \rangle = \prod_i \gamma(n_i) |\Phi_0 \rangle$,
where $|\Phi_0 \rangle$ is the non-interacting ground state and $\gamma(n_i)$ 
is an operator which progressively suppresses expensive occupancies, is
not appropriate to describe the MIT in both fermionic and bosonic 
cases.~\cite{shiba1,kotliar} 
Indeed, the only way to produce an insulating wave function corresponds to project out 
completely on-site occupancies different from the average one. This wave function, with no
charge fluctuations, is clearly a very poor description of a realistic Mott insulator. 
In fact, for the fermionic Hubbard model, the optimal $|\Psi_G \rangle$ is never insulating,
except at $U=\infty$, even if the variational wave function is improved by adding 
short-range density-density correlations.~\cite{shiba1,shiba2}
In the case of $S=0$ bosons, an insulating 
$|\Psi_G \rangle$ can be stabilized at finite $U$,~\cite{kotliar} but, as we mentioned, 
the insulator obtained in this way gives an incorrect description of the actual ground state.

\begin{figure}
\includegraphics[width=0.41\textwidth]{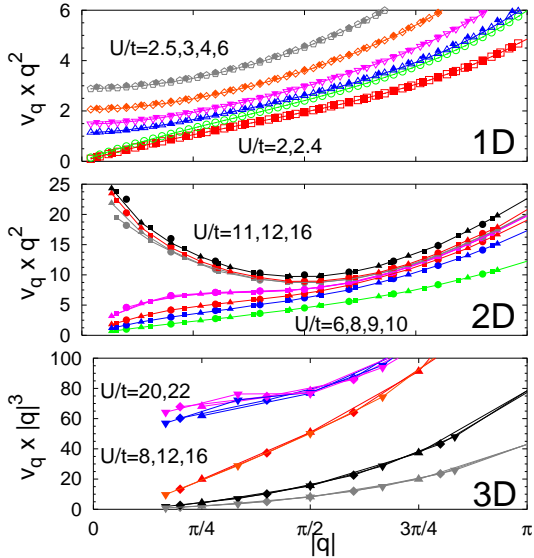}
\caption{\label{fig:jas}
Variational results for the Jastrow potential 
$v_q$ multiplied by $q^2$ in 1D and 2D and by $|q|^3$ in 3D for increasing values of $U/t$
(from bottom to top). Upper panel: 1D case for $60$ and $100$ sites. 
Middle panel: 2D case for $20 \times 20$, $26 \times 26$, and $30 \times 30$ clusters
(along the $(1,0)$ direction). Lower panel: 3D case for $8 \times 8 \times 8$, 
$10 \times 10 \times 10$, and $12 \times 12 \times 12$ clusters (along the $(1,0,0)$ direction).}
\end{figure}

\begin{figure}
\includegraphics[width=0.41\textwidth]{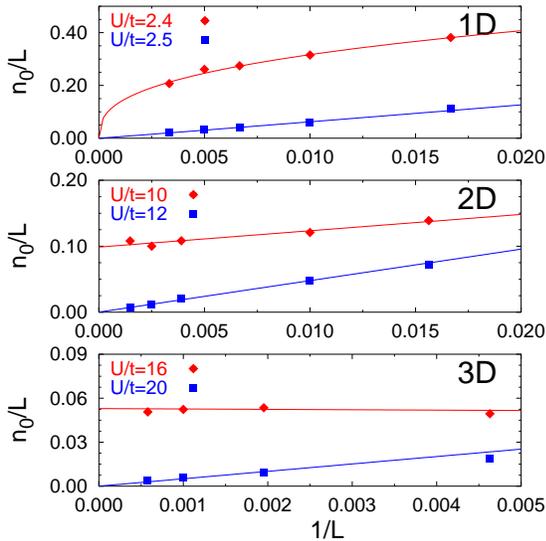}
\caption{\label{fig:nk}
Variational values of the condensate fraction ${\tt n_0}/L$ in 1D (upper panel),
2D (middle panel), and 3D (lower panel).}
\end{figure}

A step forward has been recently accomplished in one dimension, where it has been 
shown~\cite{capello} that 
a Gutzwiller wave function supplemented by a {\it long-range} Jastrow factor offers a very accurate
description of a Mott insulator. However, a systematic analysis of this variational 
ansatz in higher dimensions is still lacking, while it would be highly desirable in view of 
more realistic applications. In this letter, we apply this variational approach to the 
$S=0$ bosonic Hubbard model~(\ref{hambose}) with nearest-neighbor hopping $t/2$ in a 
one-dimensional chain (1D), a two-dimensional (2D) square lattice and a three-dimensional 
(3D) cubic lattice with $L$ sites. Specifically, we consider the following ansatz for the 
variational wave function
\begin{equation}\label{wavefunction}
|\Psi \rangle = \exp \left ( -\frac{1}{2} \sum_{i,j} v_{i,j} n_i n_j 
+g_{MB} \sum_i \xi_i \right ) |\Phi_0 \rangle,
\end{equation}
where $|\Phi_0 \rangle$ is the non-interacting fully-condensed wave function, 
i.e. $|\Phi_0 \rangle = (b_{k=0}^\dag)^N|0 \rangle$, being $b^\dagger_k$ the creation 
operator at momentum $k$ and $N$ the number of particles. In the following, we will consider
$N=L$. The components of the Jastrow potential, $v_{i,j}= v(|R_i-R_j|)$, 
are independently optimized by minimizing the variational energy,~\cite{sorella}
and we will denote by $n_q$ and $v_q$ the Fourier transforms of the local density 
$n_i$ and of $v_{i,j}$, respectively. 
Finally, $g_{MB}$ is a variational parameter related to 
the many-body operator
$\xi_i=h_i \prod_{\delta} (1-d_{i+\delta})+d_i \prod_{\delta} (1-h_{i+\delta})$,
where $h_i=1$ ($d_i=1$) if the site $i$ is empty (doubly occupied) and $0$
otherwise, and $\delta$ is the vector which connects nearest-neighbor sites;~\cite{kaplan} 
in other words, $\sum_i \xi_i$ counts the number of isolated empty and doubly occupied sites.
This term is kept mainly to improve the variational accuracy (in 2D and 3D) but does 
not introduce important correlation effects, that are instead contained only in the 
{\it long-range} behavior of the two-body Jastrow potential $v_{i,j}$. 
Remarkably, it turns out that our wave function~(\ref{wavefunction}) is quite accurate 
in all cases that we considered, even across the MIT; more details will be reported 
elsewhere.~\cite{capellobose} 

Let us start by discussing the relevance of the Jastrow factor for the low-energy properties
and some expected asymptotic behaviors of $v_q$. In the gapless (superfluid) phase, a long-range 
Jastrow potential is surely needed to restore the correct small-$q$ behavior of the static 
density structure factor, i.e., 
$N_q=\langle \Psi | n_{-q} n_q |\Psi \rangle / \langle \Psi|\Psi \rangle \sim |q|$. 
Indeed, since at least in the weak-coupling regime, the expression
\begin{equation}\label{Reatto&Chester}
N_q = \frac{\displaystyle N^0_q}{\displaystyle 1 + \gamma \,v_q\,N^0_q}
\end{equation}
holds with $\gamma=2$,~\cite{R&C} where the non-interacting 
$N_q^0=\langle \Phi_0 | n_{-q} n_q |\Phi_0 \rangle 
/ \langle \Phi_0|\Phi_0 \rangle \sim {\rm const}$, it follows that $v_q\sim 1/|q|$. 
The same asymptotic behavior was obtained in Refs.~\cite{capello,capello2,capello3} by full 
optimization of $v_q$ in metallic fermionic models both in 1D and 2D, where
both $N_q\sim |q|$ and $N^0_q\sim |q|$.
Moreover, it was shown that, in the insulating phase, a more singular $v_q\sim 1/q^2$ 
at small $q$ is required to recover the appropriate $N_q \sim q^2$ insulating behavior, 
consequence of exponentially decaying correlation functions. 
This similarity between 1D and 2D fermionic models was suggestive that 
$v_q \sim 1/q^2$ is sufficient to induce an insulating behavior in any dimensions as well 
as that the expression~(\ref{Reatto&Chester}) remains asymptotically valid 
for $|q| \to 0$, even inside the insulating regime at strong coupling.
However, one easily realizes that, were this conclusion correct, the 
variational wave function~(\ref{wavefunction}) could not describe 
any bosonic insulator in 3D, 
since $v_q \sim 1/q^2$ is not sufficient to empty the condensate fraction 
${\tt n_0} = \langle \Psi| b^\dagger_{k=0}\,b^{\phantom{\dagger}}_{k=0}|\Psi\rangle
/\langle \Psi|\Psi\rangle$.~\cite{reatto} 
Instead, we will show that our Jastrow wave function can give a consistent description of 
the Mott phase {\it in any dimension} thanks to an even more diverging $v_q$, implying that  
the formula~(\ref{Reatto&Chester}) can be violated in the real 3D world.

In Fig.~\ref{fig:jas} we present the optimized Jastrow potential $v_q$. For all
dimensions, the MIT can be clearly detected from the sudden change in the small-$q$ 
behavior of $v_q$. On the one hand, the gapless superfluid phase is always described by 
$v_q \sim \alpha/|q|$, with $\alpha$ increasing with $U$. On the other hand, 
the gapped insulator has a much more diverging $v_q$. In 1D we recover the 
$v_q \sim 1/q^2$ behavior, like in the fermionic case.~\cite{capello}
In 2D, the leading behavior of the Jastrow potential across the transition is less 
clearcut than in 1D. Indeed, we cannot establish whether, on the insulating side, 
$v_q\sim \beta_{2D}/q^2$ with $\beta_{2D}$ large but finite, or possible logarithmic 
corrections have to be considered, i.e., $v_q \sim \ln(1/|q|)/q^2$.
The first possibility is particularly appealing since, in this case, the insulating 
phase can be interpreted in terms of the confined phase of the 2D classical 
Coulomb gas.~\cite{capello3,minnhagen}
Notice that the optimized $v_q$, that also contains subleading corrections to the $1/q^2$
behavior, can modify the critical properties of the classical Coulomb gas model. 
Nevertheless, the essential point is that, within this approach, 
the MIT can be still interpreted in terms 
of a binding-unbinding transition among charged particles (empty and doubly occupied sites).
Finally, in 3D an even more diverging $v_q$ is stabilized in the insulating regime, i.e.,
$v_q \sim 1/|q|^3$. Therefore, in all these cases the Jastrow potential is sufficient to 
destroy the condensate (see Fig.~\ref{fig:nk}).~\cite{notenk}

\begin{figure}
\includegraphics[width=0.41\textwidth]{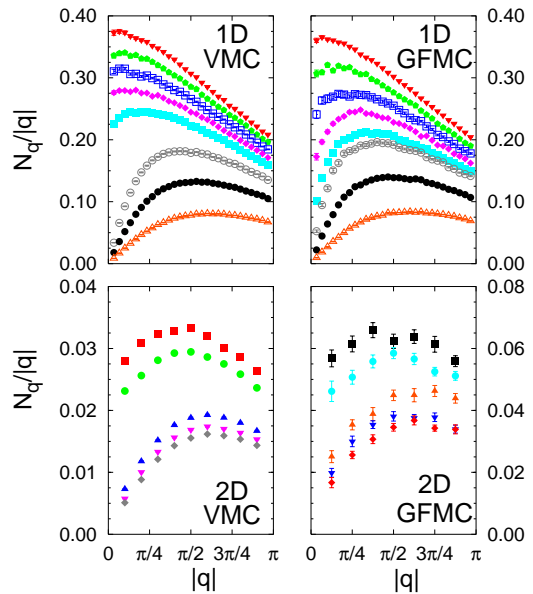}
\caption{\label{fig:nq1d2d}
Density structure factor $N_q$ divided by $|q|$ calculated with variational
Monte Carlo (left panels) and GFMC (right panels) in 1D (upper panels) and 2D (lower panels)
In 1D, $L=60$ and $U/t=1.6$, $1.8$, $2$, $2.2$, $2.4$, $2.5$, $3$, and $4$.
In 2D, $L=20 \times 20$ and $U/t=10$, $10.2$, $10.4$, $10.6$, and $10.8$ for the 
variational calculation, and $L=256$ and $U/t=8$, $8.2$, $8.4$, $8.6$, and $8.8$ in the 
GFMC calculation. All cases are shown from top to bottom for increasing values of $U/t$.}
\end{figure}

In order to verify the validity of our approach, let us move to discuss the variational 
results for the density structure factor $N_q$ in comparison 
with the exact ones obtained by GFMC. For small $q$'s we can generally write 
$N_q=\gamma_1 |q| + \gamma_2 q^2 + \mathrm{O}(q^3)$.
In analogy with spin systems, we may assume that $\gamma_1 = v_c \, \chi$, being $v_c$
and $\chi$ the charge  velocity and the compressibility, respectively. 
At the variational level $\gamma_1$ and $\gamma_2$ depend crucially upon the Jastrow 
parameters. Indeed, we do find that, in the superfluid phase, $\gamma_1 \ne 0$ while, 
in the Mott insulator, $\gamma_1=0$ and $\gamma_2 \not = 0$, signaling that this state 
is incompressible (see Fig.~\ref{fig:nq1d2d}). 
Moreover, in 1D we have evidence for a jump (from a finite value to zero) in $\gamma_1$ 
across the MIT, especially because its value does not change much passing from $U=U_c/2$
and $U \lesssim U_c$ (e.g., it changes from $0.4$ to $0.2$).
These variational results are confirmed by GFMC and are consistent with the
finite jump of the compressibility across the Kosterlitz-Thouless 
transition, expected in 1D.~\cite{fisher}
Our numerical results seem also to indicate that $\gamma_2$ diverges as the MIT is 
approached from the insulating side. In the variational calculation, this behavior follows 
from a $v_q \sim \beta_{1D}/q^2$ in the insulating phase with $\beta_{1D} \to 0$ at the MIT.
In conclusion we find that the 1D MIT can be located at $U_c/t \simeq 2.45$ 
in the variational calculations, whereas the GFMC gives $U_c/t \simeq 2.2$
(in close agreement with previous calculations of Ref.~\cite{batrouni,kuhner}), 
showing that the variational wave function~(\ref{wavefunction}) is not only 
qualitatively but also quantitatively correct. 

\begin{figure}
\includegraphics[width=0.41\textwidth]{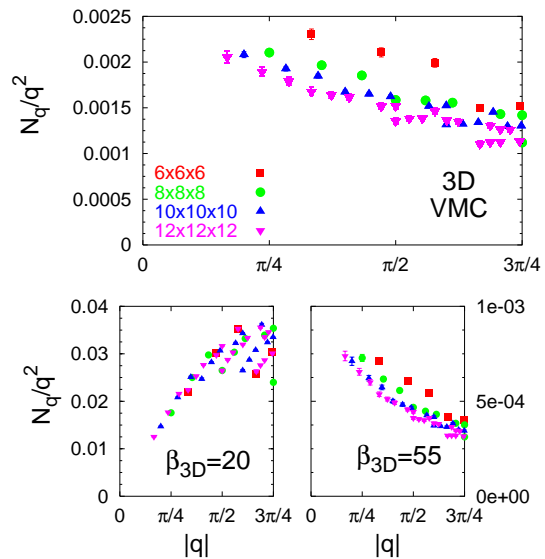}
\caption{\label{fig:nq3d}
Upper panel: Density structure factor $N_q$ divided by $q^2$ calculated by the
variational Monte Carlo for 3D and $U/t=20$.
Lower panels: $N_q$ for non-optimized wave functions with $v_q \sim \beta_{3D}/|q|^3$ for
two values of $\beta_{3D}$ and the same sizes as above.}
\end{figure}

The density structure factor $N_q$ displays quite distinct long-wavelength 
behaviors for weak and strong interactions also in 2D, see Fig.~\ref{fig:nq1d2d}.
In the variational calculations, for $U/t \lesssim 10.3$ the structure factor
goes like $N_q \sim \gamma_1 |q|$, while for $U/t \gtrsim 10.3$ we get
$N_q \sim \gamma_2 q^2$. The critical value of the on-site interaction is
slightly different from the GFMC one, which we find to be $U_c/t \simeq 8.5$, in
agreement with Ref.~\cite{krauth}.
In spite of the different values of $U_c$, the qualitative behavior across the MIT 
is similar both in the variational and in the GFMC calculations.
It should be emphasized that in 2D the value of $\gamma_1$ close to the MIT is
one order of magnitude smaller than its value at $U/U_c \sim 1/2$, 
a behavior qualitatively different from 1D, suggesting  that $\gamma_1$ vanishes 
upon approaching the MIT from the superfluid side and that $\gamma_2$ is smooth 
and constant across the MIT in 2D.

More interesting is the 3D case. Here, the GFMC is severely limited by small sizes and,
therefore, we will just discuss the variational results. 
The change in the leading behavior of the Jastrow parameters $v_q$ allows us to
locate the transition around $U_c/t \simeq 18$, which is very close to the
critical value extracted from experiments on optical lattices.~\cite{greiner}
The optimal Jastrow potential 
behaves as usual as $v_q\sim \alpha/|q|$ in the superfluid phase, 
but turns into 
$v_q\sim \beta_{3D}/|q|^3$ in the Mott insulator (see Fig.~\ref{fig:jas}). This behavior 
would imply, if Eq.~(\ref{Reatto&Chester}) were valid, a charge structure factor 
$N_q \sim |q|^3$. By contrast, we do find that $N_q \sim q^2$, as expected in an insulator,
see Fig.~\ref{fig:nq3d}. So, we arrive at the very surprising and unexpected result 
that Eq.~(\ref{Reatto&Chester}) does not hold, not even asymptotically for $|q|\to 0$. 
In order to prove more firmly that a $v_q\sim \beta_{3D}/|q|^3$ can indeed lead to 
$N_q\sim q^2$, we have calculated $N_q$ with a non-optimized wave function of the 
form~(\ref{wavefunction}) with $v_q\sim \beta_{3D}/|q|^3$, for different values of 
$\beta_{3D}$.
As shown in Fig.~\ref{fig:nq3d}, for small $\beta_{3D}$'s $N_q\sim |q|^3$, implying that 
Eq.~(\ref{Reatto&Chester}) is qualitatively correct. However, above a critical 
$\beta^*_{3D}$, the behavior turns into $N_q \sim q^2$, signaling a remarkable breakdown 
of Eq.~(\ref{Reatto&Chester}).
The optimal value of $\beta_{3D}$ that we get variationally at the MIT is larger than 
$\beta^*_{3D}$, confirming our variational finding $N_q \sim q^2$. We notice that the 
change of behavior as a function of $\beta_{3D}$ is consistent with the binding-unbinding 
phase transition recently uncovered in a classical 3D gas with potential 
$V_{\rm cl}(q) \sim 1/|q|^3$.~\cite{3dcgm} 

In conclusion we have demonstrated that a long-range Jastrow potential does allow for a 
faithful variational description of a Mott transition in the bosonic Hubbard model, 
in spite of the fact that the uncorrelated wave function onto which the Jastrow factor 
is applied has full Bose-condensation. An interesting outcome of our analysis is that, 
in 3D, the Mott insulator is characterized by a very singular Jastrow potential, 
$v_q \sim 1/|q|^3$, that is able to empty the condensate, yet leading to a well behaved 
charge structure factor, $N_q\sim q^2$. This result contradicts the na\"{\i}ve expectation,
$N_q\sim 1/v_q$, based on the weak-coupling formula~(\ref{Reatto&Chester}).  
This breakdown of the weak-coupling approach is the necessary condition for our wave function 
to work in 3D and represents a highly nontrivial consistency check of our non-perturbative 
variational theory of the Mott phase. We argue that this variational theory will hold 
also in electronic models. In particular, once the square of the ground state wave function 
is interpreted as a classical partition function, the metal-insulator transition can be 
induced in any dimension by a singular interaction between charge fluctuations. Remarkably,
in $D>1$ this interaction remains always logarithmic, suggesting an unconventional 
binding-unbinding description of the metal-insulator transition.
In analogy with the bosonic example we have analyzed, we should expect that a singular 
Jastrow potential, $v_q\sim 1/|q|^\theta$ with $\theta= 3$, might be necessary to describe 
the 3D Mott transition in fermionic models, too, all the more reason when realistic 
Coulomb interaction is taken into account.

We thank important discussions with D. Poilblanc and T. Senthil.
This work has been partially supported by CNR-INFM and
COFIN 2004 and 2005.


\begin{thebibliography}{99}
\bibitem{notenk} In 1D, even $v_q \sim 1/|q|$ is sufficient to completely suppress the
   condensate fraction, as expected from Ref.~\cite{reatto}.
\end{thebibliography}
\end{document}